\documentclass[aps,twocolumn,prl,showpacs,preprintnumbers]{revtex4}
\renewcommand\d{\partial}

\usepackage{amsmath,amssymb,bm,comment,color}
\usepackage{hyperref}

\newcommand{\ddd}{\displaystyle}
\newcommand{\lbr}{\left(}
\newcommand{\rbr}{\right)}

\newcommand{\CS}{S}

\def\d{\partial}

\def\nn{\nonumber}

\def\ttau{\tilde \tau}

\advance\voffset by 0.5cm

\begin{document}

\preprint{DESY 11-022}

\title{Fluid-gravity model for the chiral magnetic effect}
\author{Tigran Kalaydzhyan}
\author{Ingo Kirsch \medskip}

\affiliation{DESY Hamburg, Theory Group, Notkestrasse 85, D-22607 Hamburg, Germany}
\date{February 2011}
\begin{abstract}
We consider the STU model as a gravity dual of a  strongly coupled plasma with multiple 
anomalous $U(1)$ currents. In the bulk we add additional background gauge fields to include 
the effects of external electric and magnetic fields on the plasma. Reducing the number
of chemical potentials in the STU model to two and interpreting them as quark and chiral
chemical potential, we obtain a holographic description of the chiral magnetic and chiral
vortical effects (CME and CVE) in relativistic heavy-ion collisions. These effects formally
appear as first-order transport coefficients in the electromagnetic current. We compute
these coefficients from our model using fluid-gravity duality. We also find analogous
effects in the axial-vector current. Finally, we briefly discuss a variant of our model, in which
the CME/CVE is realized in the late-time dynamics of an expanding plasma.
\end{abstract}
\maketitle

\section{I. Introduction}

In the past two years, the chiral magnetic effect (CME) \cite{Kharzeev, Fukushima}
has received much attention in lattice QCD \cite{lattice}, hydrodynamics
\cite{Isachenkov}-\cite{Zakharov} and holographic models \cite{Lifschytz}-\cite{Landsteiner2}.
The CME states that, in the presence of a magnetic field ${\bf B}$, an electric
current of the type
\begin{align}
 {\bf J} =  C \mu_5 {\bf B} \,,\qquad C=\frac{N_c}{2\pi^2} 
\end{align}
is generated in the background of topologically non\-trivial gluon fields.
This could possibly explain the charge asymmetry observed in heavy-ion collisions 
at the BNL Relativistic Heavy Ion Collider \cite{STAR}.

A hydrodynamic description of the CME has recently been found in \cite{Isachenkov, Pu},
using techniques developed in \cite{Son}. This model contains two $U(1)$ currents, 
an axial and a vector one, which are assumed to be conserved, at least in the absence
of electric fields. This allows us to introduce the corresponding chemical potentials 
$\mu$ and $\mu_5$, and the CME was shown to arise as a first-order transport coefficient
$\kappa_B$ in the constitutive equation for the 
electromagnetic current $\Delta j^\mu = \kappa_B B^\mu$ with 
$\kappa_B = C\mu_5$. 

There have also been several proposals for a holographic description
of the CME. In the early works \cite{Lifschytz, Yee_CME}, the
axial anomaly was not realized in covariant form and the 
electromagnetic current was not strictly conserved. 
Aiming at restoring the conservation of the electromagnetic 
current, Ref.~\cite{Rebhan} introduced the Bardeen counter\-term
into the action. However, as shown in \cite{Rubakov, Rebhan} for some
standard AdS/QCD models, this typically leads to a vanishing
electromagnetic current.

In \cite{Rubakov, Landsteiner} the problem was traced back to the difficulty of introducing
a chemical potential conjugated to a nonconserved charge.
Reference~\cite{Rubakov} therefore suggested a modification of the action
in which the axial charge is conserved. This
charge is, however, only gauge invariant when integrated
over all space in homogeneous configurations \cite{Landsteiner}, while the 
charge separation in heavy-ion collisions is clearly inhomogeneous.
In contrast, Ref.~\cite{Landsteiner} introduced a chiral chemical potential dual to
a gauge invariant current, despite it being anomalous. 
This required a singular bulk gauge field at the horizon, a phenomenon 
which seems to be generic in AdS black hole models of the CME.

In this Letter, we propose a different approach which is based on 
the fluid-gravity correspondence \cite{Hubeny} rather than a static (AdS/QCD) model.
The fluid-gravity duality is more flexible since the hydrodynamic gradient expansion 
captures (small) deviations from equilibrium. This includes the CME
and the change of the chiral charge density due to the anomaly $E \cdot B$ as first-
and second-order effects, respectively. This allows us to introduce 
chemical potentials even for anomalous currents.

Our main goal is to construct a holographic dual of the
hydrodynamic two-charge model of Ref.~\cite{Isachenkov}. We will start
from the three-charge STU model \cite{Cvetic} which we take as a prototype of an AdS
black hole background with several $U(1)$ charges. We consider it as a
 {\em phenomenological} model of a strongly coupled plasma
with multiple chemical potentials; {\em i.e.},\ we prescind from the
strict string theory interpretation of the three $U(1)$'s as 
 R charges inside the $SO(6)_R$ R symmetry of ${\cal N}=4$ super-Yang-Mills
theory. This allows us to interpret one of them as an axial $U(1)$ charge and the other
two as a single vector $U(1)$ charge. These charges are dual to
$\mu$ and $\mu_5$ required in the hydrodynamic description \cite{Isachenkov}. 

We proceed as follows.
First, we show that the two-charge model of  \cite{Isachenkov} can be considered
as a special case of the hydrodynamic three-charge  model of~\cite{Son, Oz}.  
Next, we reproduce the relevant magnetic conductivities in this three-charge model 
from the dual STU model (plus background gauge fields),
using fluid-gravity duality \cite{Hubeny}. 
We then reduce the model to two charges and recover the CME ({\em i.e.}\ $\kappa_B$ 
of \cite{Isachenkov}) as well as other related effects. 
Finally, we present a time-dependent version of
the STU model dual to a boost-invariant expanding plasma.

\section{II. CME and CVE in hydrodynamics}

{\em Hydrodynamics of a $U(1)^3$ plasma.} The hydrodynamic regime
of relativistic quantum gauge theories with
triangle anomalies has been studied in \cite{Son, Oz}. 
The anomaly coefficients are usually given by a totally symmetric
\mbox{rank-$3$} tensor $C^{abc}$ and the hydrodynamic equations  
are
\begin{align}\label{hydroeqn}
  \partial_\mu T^{\mu\nu} = F^{a\nu\lambda} j^{a}_\lambda
  \,,\qquad
  \d_\mu j^{a\mu} = C^{abc} E^b\cdot B^c \,,
\end{align}
where $E^{a\mu}=F^{a \mu\nu} u_\nu$, $B^{a\mu}=\frac{1}{2}\epsilon^{\mu\nu\alpha
\beta}u_\nu F^a_{\alpha\beta}$ ($a=1,2,3$) are electric and magnetic fields, 
and $F^a_{\mu\nu}=\partial_\mu A^a_\nu-\partial_\nu A^a_\mu$ denotes the gauge
field strengths. The stress-energy
tensor $T^{\mu\nu}$ and $U(1)$ currents $j^{a\mu}$  are
\begin{align}
T^{\mu\nu}&= (\epsilon+P) u^\mu u^\nu + P g^{\mu\nu} + \tau^{\mu\nu} \,,\nonumber\\
j^{a\mu} &= \rho^a u^\mu + \nu^{a \mu}\,, \label{currents}
\end{align}
where $\tau^{\mu\nu}$ and $\nu^{a \mu}$ denote higher-gradient corrections.
$\rho^a$, $\epsilon$, and $P$ denote the charge densities, energy density and pressure,
respectively.

In the presence of $E$ and $B$ fields the first-order correction of the $U(1)$
currents is given by
\begin{equation}
  \nu^{a\mu} = \xi^a_\omega \omega^\mu + \xi_B^{ab} B^{b\mu} + ...\,,
\end{equation}
where $\omega^\mu\equiv \frac{1}{2} \epsilon^{\mu\nu\lambda\rho} u_\nu\partial_\lambda u_\rho$ is the
vorticity. The ellipses indicate further terms involving electric fields.
The conductivities $\xi^a_\omega$ and $\xi_B^{ab}$ were first introduced
in \cite{Erdmenger, Son} and are given by \cite{Oz} (see also \cite{Son, Zahed}) 
\begin{align}
  \xi^a_\omega &= C^{abc} \mu^b\mu^c 
    - \frac{2}{3} \rho^a C^{bcd} \frac{\mu^b\mu^c\mu^d}{\epsilon+P}\,,\label{xi}\\
 \xi_B^{ab} &=  C^{abc} \mu^c 
    - \frac{1}{2}\rho^a C^{bcd} \frac{\mu^c\mu^d}{\epsilon+P} \,,\label{xiB}
\end{align}
where $\mu^a$ are the three chemical potentials.
These conductivities are specific for relativistic quantum field theories
with quantum anomalies and do not appear in nonrelativistic theories \cite{Son}.

\medskip
{\em Magnetic and vortical Effects. }
For the hydrodynamical description of the chiral magnetic effect, we only need
an axial and a vector chemical potential, $\mu_5$ and~$\mu$.
The three-charge model can be reduced to one with two charges by 
choosing the following identifications:
\begin{align} \label{identis}
 {A}_\mu^A &= {A}^1_\mu \,,\qquad {A}_\mu^V = {A}^2_\mu
= { A}^3_\mu \,,\nonumber\\
 \mu_5 &=\mu^1 \,, \qquad \mu=\mu^2=\mu^3 \,, \nonumber\\
 j^{\mu}_5&=j^{1\mu}  \,, \qquad j^{\mu}= j^{2\mu}+j^{3\mu} \,,
\end{align}
and $C^{123}=C^{(123)}=\frac{C}{2}$. In the absence of axial gauge fields $A^A_\mu$ 
(which are not required), (\ref{hydroeqn}) simplifies to
\begin{align} \label{hydroeqn2}
  \partial_\mu T^{\mu\nu} \simeq  F^{V \nu\lambda} j_\lambda  
  \,,\quad
  \d_\mu j^{\mu}_5 = C  E^\lambda  B_\lambda \,,\quad
  \d_\mu j^{\mu} \simeq 0 \,,
\end{align}
where $E^{\mu}\equiv F^{V \mu\nu} u_\nu$, $B^{\mu} \equiv\frac{1}{2}\epsilon^{\mu\nu\alpha
\beta}u_\nu F^V_{\alpha\beta}$. The symbol ``$\simeq$'' indicates that the equation only holds
for $A^A_\mu=0$. 

Let us also define
\begin{align}
\rho_5&=\rho^1 \,,\quad \rho=\rho^2+\rho^3\,, \nn\\
\kappa_\omega&=\xi^2_\omega+\xi^3_\omega\,,\quad \kappa_B =\xi_B^{22}+\xi_B^{23}+\xi_B^{32}+\xi_B^{33}
\,,\nn\\
\xi_\omega&=\xi^1_\omega\,,\quad  
\xi_B =\xi_B^{12}+\xi_B^{13}   \,. \label{idxi}
\end{align}
Then from (\ref{currents})--(\ref{xiB}) we get the constitutive equations 
\begin{align}
j^\mu & = \rho u^\mu + \kappa_\omega\omega^\mu + \kappa_B B^{\mu}\,, \nonumber\\
j^{\mu}_5 & = \rho_5 u^\mu + \xi_\omega\omega^\mu + \xi_B B^{\mu}\,,
\end{align}
with coefficients
\begin{align}
\!\kappa_\omega&= 2 C \mu\mu_5 \lbr 1 - \ddd\frac{\mu \rho}{\epsilon + P} \rbr, \quad \kappa_B = C \mu_5 \lbr 1 - \ddd\frac{\mu \rho}{\epsilon + P} \rbr ,\nonumber\\
\!\xi_\omega&= C \mu^2 \lbr 1 - 2\ddd\frac{\mu_5 \rho_5}{\epsilon + P} \rbr\!,\,\,\, \xi_B = C \mu \lbr 1 - \ddd\frac{\mu_5 \rho_5}{\epsilon + P} \rbr\! , \label{effects}
\end{align}
which to leading order are in agreement with \cite{Isachenkov}. 

The leading term in $\kappa_B$ is nothing but the {\em chiral magnetic effect} (CME)
\cite{Kharzeev, Fukushima}, $\kappa_B = C\mu_5$. There is a second effect given by the term
$\kappa_\omega=2C\mu\mu_5$ which has recently been termed
{\em chiral vortical effect} (CVE) \cite{CVE}. The CVE states that, if the liquid rotates with 
some angular velocity $\vec \omega$, an electromagnetic current is 
induced along $\vec \omega$ -- there are analogous effects in the axial current $j^\mu_5$. The leading
term in $\xi_B$,  $\xi_B = C \mu$, generates an axial current parallel to the magnetic field.  
There is also a vortical effect given by $\xi_\omega= C \mu^2$. 	
This term describes chirality separation through rotation \cite{Zakharov}.
We may refer to these effects as quark magnetic (QME) and quark vortical effects (QVE)
since $\xi_\omega/\mu$ and $\xi_B$ are proportional to the quark chemical potential
$\mu$ (while $\kappa_\omega/\mu$ and $\kappa_B \propto \mu_5$).

We may also shift all anomalies in (\ref{hydroeqn}) entirely into the current
$j^{1\mu}$ $(=j^\mu_5)$ by adding Bardeen currents,
\begin{align}
j'{}^{\mu} &\equiv  j^\mu + j_B^\mu\,, \qquad j'{}^{\mu}_{\!\!5\,\,} \equiv j^\mu_5 + j_{5,B}^\mu\,,
\nonumber\\
j^\mu_B &= c_B \varepsilon^{\mu\nu\lambda\rho} (A^V_\nu F_{\lambda\rho}^A 
- 2 A_\nu^A F^V_{\lambda\rho} )\,, \nonumber \\
j^\mu_{5,B} &= c_B \varepsilon^{\mu\nu\lambda\rho} A^V_\nu F^V_{\lambda\rho} \,,
\label{Bardeen}
\end{align}
with $c_B=-C/2$ such that (\ref{hydroeqn}) becomes ($C'=3C$)
\begin{align}
  \partial_\mu T^{\mu\nu} \simeq  F^{V \nu\lambda} j'_\lambda  
  \,,\quad
  \d_\mu j'{}^{\mu}_{\!\!5\,\,} = C'  E^\lambda  B_\lambda \,,\quad
  \d_\mu j'{}^{\mu} = 0 \,. \nonumber
\end{align}
This is formally identical to (\ref{hydroeqn2}), leading again to~(\ref{effects}). 

\section{III. Fluid-gravity model for the CME}

{\em Three-charge STU model with external fields.}
In the following we propose the three-charge STU model~\cite{Cvetic}
as a holographic dual gravity theory for the (chiral) magnetic and vortical effects
in a relativistic fluid. We begin by showing that the first-order transport
coefficients (\ref{xi}) and ({\ref{xiB}) of the $U(1)^3$ theory can be reproduced
from the STU model~\cite{Cvetic}. Subsequently, we will reduce it
to a two-charge model and recover the conductivities (\ref{effects}). 

The Lagrangian of the STU-model is given by~\cite{Cvetic}
\begin{align}
{\cal L}&= R\,-\,{\frac{1}{2}}G_{ab}F^a_{MN} F^{bMN}
-G_{ab}\partial_M X^a \partial^M X^b + 4\sum_{a=1}^3 {1\over X^a}  \nn\\
&~~~+ {\frac{1}{24}\sqrt{-g_5}}\epsilon^{MNPQR} \CS_{abc}
F^a_{MN}F^b_{PQ}A^c_R \,,\label{Lagrangian}
\end{align}
where 
\begin{align}
 G_{ab}={1\over 2}\delta_{abc}(X^c)^{-2},\qquad X^1 X^2 X^3 =1 \,.
\end{align}
$g_{MN}$, $X^a$ and $A^a_M$ ($M,N=0,1,...,4$, $a,b,c=1,2,3$) denote the metric, three scalars and $U(1)$
gauge fields, respectively. 

The boosted black brane solution corresponding to the three-charge STU model
is given by \cite{Cvetic}
\begin{align}
ds^2 &= -H^{-{2\over 3}}(r)f(r) u_\mu u_\nu dx^\mu dx^\nu
-2 H^{-{1\over 6}}(r) u_\mu dx^\mu dr \nonumber \\
&~~~+ r^2 H^{1\over 3}(r)  \left(\eta_{\mu\nu}+u_\mu u_\nu\right)dx^\mu dx^\nu \,,
\nonumber\\
A^a &= \left(A_0^a(r) \,u_\mu + {\cal A}^a_\mu \right)  dx^\mu \,,\quad
X^a \,\,=\,\, {H^{1\over 3}(r) \over H_a(r)}\,, \label{0thsol}
\end{align}
where ($\mu,\nu=0,1,2,3$)
\begin{align}
f(r)&=-{m\over r^2}+r^2 H(r)\,,\quad H(r)=\prod_{a=1}^3 H^a(r)\,,\nonumber\\
 H^a(r)&=1+{q^a \over r^2}\,,\qquad A_0^a(r) = {\sqrt{m q^a}\over r^2+q^a}  \label{mu}\,,
\end{align}
and $u_\mu$ is the four-velocity of the fluid with $u_\mu u^\mu=-1$. Following \cite{Son},
we have formally introduced constant background gauge fields ${\cal A}^a_\mu$, which are
necessary for the computation of the transport coefficients $\xi_B^{ab}$. 
\medskip

We now use the standard procedure \cite{Hubeny} to holographically compute the transport
coefficients $\xi^a_\omega$ and $\xi^{ab}_B$. We closely follow \cite{Yee} which has already
determined $\xi^a_\omega$ from the STU-model (but not $\xi^{ab}_B$ relevant for the CME).
Working in the frame $u_\mu=(-1,0,0,0)$ (at $x^\mu=0$),
we slowly vary $u_{\mu}$ and ${\cal A}^a_\mu$ 
up to first order as
\begin{align}
 u_{\mu} = (-1, x^{\nu}\partial_{\nu}u_i),\qquad {\cal A}^a_\mu = (0, x^{\nu}\partial_{\nu} {\cal A}^a_i) \,.\label{variations}
\end{align} 
We may also vary $m$ and~$q$ in this way, but it turns out that 
varying these parameters has no influence on the transport
coefficients $\xi^a_\omega$ and $\xi_B^{ab}$.

As a consequence, the background (\ref{0thsol}) is no longer an
exact solution of the equations of motion but receives higher-order corrections. 
The corrected  metric and gauge fields can  be rewritten in Fefferman-Graham coordinates
and expanded near the boundary (at $z=0$) as 
\begin{align}
ds^2 &= \frac{1}{z^2}\left(g_{\mu\nu}(z,\,x)\, dx^\mu dx^\nu + dz^2\right)\,,\nonumber\\
&g_{\mu\nu}(z, x) = \eta_{\mu\nu} + g^{(2)}_{\mu\nu}(x)\, z^2 
+ g^{(4)}_{\mu\nu}(x)\, z^4 + ... \,,\nonumber \\
 &A^a_{\mu}(z,\,x) = A^{a (0)}_{\mu}(x) +  A^{a (2)}_{\mu}(x)\, z^2 + ...\,.
\end{align}

The first-order gradient corrections of the energy-momentum tensor and 
$U(1)$ currents (\ref{currents}) are then read off from \cite{Skenderis, Yee3}
\begin{align}
T_{\mu\nu} &= \frac{g^{(4)}_{\mu\nu}(x)}{4\pi G_5} + c.t.\,,\quad  \label{J} 
j_a^\mu = \frac{\eta^{\mu\nu}A_{a\nu}^{(2)}(x)}{8\pi G_5}  + \hat j_a^\mu \,, \\
&\hat j_a^\mu = - \frac{\CS_{abc}}{32\pi G_5}
\epsilon^{\mu\nu\rho\sigma}A_{b \nu}^{(0)}(x)\partial_\rho A_{c \sigma}^{(0)}(x) 
\label{jhat}
\,,
\end{align}
where $c.t.$~denotes diagonal corrections to the energy-momentum tensor due to counterterms.
The term $\hat j_a^\mu$ will be discussed below around (\ref{furtherterm}).

The computation of the corrected metric and gauge fields is very similar to that
in \cite{Yee}. At zeroth order, we get the same expressions
for the pressure $P$ and charge densities $\rho_a$
as in \cite{Yee}, $P\equiv {m}/{16\pi G_5}$ and $\rho_a\equiv {\sqrt{m q_a}}/{8\pi G_5}$,
which may be combined to give 
\begin{align}
\label{replacement}
{\sqrt{m q^a}\over 2m} = \frac{\rho^a}{\epsilon + P} \qquad (\epsilon=3P)\,.
\end{align}
At first order, the transport coefficients $\xi^a_\omega$ and $\xi^{ab}_B$ are read off
from the near boundary behavior of $A^a_{\mu}$ via (\ref{J}),
\begin{align}
 \xi^a_\omega&= {1\over 16 \pi G_5}\Bigg( \CS^{abc} {\mu^b\mu^c}
- {\sqrt{m q^a}\over 3m} \CS^{bcd} \mu^b\mu^c\mu^d \Bigg) \label{xi_hol}\,,\\
\xi^{ab}_B &= {1\over 16 \pi G_5}\Bigg( \CS^{abc} {\mu^c}
- {\sqrt{m q^a}\over 4m} \CS^{bcd} \mu^c\mu^d \Bigg)\,,\label{xiB_hol}
\end{align}
with $\mu^a\equiv A^a_0(r_H) - A^a_0(\infty)$. Using a standard relation between
the anomaly coefficients $C_{abc}$ and the Chern-Simons parameters $\CS_{abc}$,
 $C_{abc} = \CS_{abc}/{16\pi G_5}$,
as well as (\ref{replacement}), we find that the holographically computed transport coefficients (\ref{xi_hol}) and (\ref{xiB_hol}) coincide exactly with those found in hydrodynamics, 
(\ref{xi}) and (\ref{xiB}).

\medskip
{\em Holographic magnetic and vortical effects.}
In order to obtain the holographic versions of the magnetic and vortical effects
(\ref{effects}), we reduce the STU model to a two-charge model using the
same identities as in hydrodynamics, (\ref{identis}) and (\ref{idxi}).
In particular, we define
(vector and axial) gauge fields ${ A}_\mu^V$ and ${ A}_\mu^A$, chemical potentials
$\mu$ and $\mu_5$, and currents $j^\mu$ and $j^\mu_5$ as in (\ref{identis})
but now with $\mu^a \equiv A^a_0(r_H) - A^a_0(\infty)$, and $A^a$ and $j_a^\mu$ as
in (\ref{0thsol}) and (\ref{J}), respectively.
Moreover, we set $\CS_{abc} = S/2$ with $S=16\pi G_5 C$ and keep
$S_{abc}$ general, as in \cite{Son}.
 
Using also the identifications (\ref{idxi}), but now for the holographically computed
transport coefficients (\ref{xi_hol}) and~(\ref{xiB_hol}), we get
\begin{align}
\kappa_\omega &= {2C \mu\mu_5}\lbr 1-\mu\sqrt{\frac{q}{m}} \rbr\,, \quad
\kappa_B= {C \mu_5}\lbr 1-\mu\sqrt{\frac{q}{m}}\rbr , \nonumber\\
\xi_\omega &= {C \mu^2} \lbr 1-\mu_5\sqrt{\frac{q_5}{m}}\rbr\,, \quad 
\xi_B={C \mu}\lbr 1-\frac{\mu_5}{2}\sqrt{\frac{q_5}{m}}\rbr ,\nonumber
\end{align}
in agreement with (\ref{effects}). This shows that the CME, CVE, etc.\ are realized 
in the STU-model, when appropriately reduced to a two-charge model. 
\medskip

{\em Comments.} To get an anomaly free three-point function for $j^\mu$, we also need
to add the Bardeen currents (\ref{Bardeen}). As in hydrodynamics, this does not change
the structure of the transport coefficients. Note however that, together
with (\ref{jhat}), the Bardeen term gives rise to additional contributions
of the type
\begin{align}
\Delta j^\mu&=\hat j^{2\mu}+\hat j^{3\mu} +j^{\mu}_B \nonumber\\
&\supset \varepsilon^{\mu\nu\rho\sigma}
({\cal A}_\nu^A(x) {\cal F}^V_{\rho\sigma}(x)- {\cal A}_\nu^V(x) {\cal F}^A_{\rho\sigma}(x)) \,.
\end{align}
If we choose ${\cal A}^A_\nu= \alpha^A u_\nu$ (at \mbox{$x=0$}) with some constant $\alpha^A\neq 0$, 
we get terms of the type ${\cal A}_0^A B^\mu$ which are forbidden
by electromagnetic gauge invariance \cite{Rubakov}. As in \cite{Rubakov, Landsteiner},
we are therefore forced to switch off the axial background
gauge field ${\cal A}_\mu^A$
completely, $\alpha^A=0$.
This corresponds to a nonvanishing gauge field at the horizon,  as in \cite{Landsteiner}.
There is also an additional term in~$j_5^\mu$, 
\begin{align} \label{furtherterm}
\Delta j^\mu_5=
\hat j^{\mu}_5+ j_{5,B}^{\mu} \propto \epsilon^{\mu\nu\rho\sigma}{\cal A}_{\nu}^{V}(x) 
{\cal F}_{ \rho\sigma}^{V}(x) \,,
\end{align}
which is of second order, as can be seen by substituting 
${\cal A}^V_{\nu}(x)=(0,x^\nu\partial_\nu {\cal A}^V_{i})$.
For instance, choosing ${\cal A}_\mu^{V} = (0,-x_2 B,0,-tE)$, {\em i.e.}\
constant E and B fields along $x^3$, the charge density $\rho_5\equiv j_5^0$ changes
linearly in time, $\Delta \rho_5 \sim S \epsilon^{0321} {\cal A}_3^V \partial_2 {\cal A}_1^V \sim t \,C E B$, as expected. Thus, holographic renormalization perfectly takes into account changes of 
hydrodynamic currents due to the anomaly, showing that fluid-gravity duality is consistent
even for nonconserved currents. [The effects enter 
via the Chern-Simons parameters $S^{abc}\propto C^{abc}$ in (\ref{jhat}).]

\medskip
{\em Holographic time-dependent model for the CME.} --
It is well known that the hydrodynamic gradient expansion 
of a fluid is also realized in
the late-time evolution of a boost-invariant expanding plasma 
{\em \`a la} \cite{Janik}. Recently, a time-dependent Reissner-Nordstr\"om-type solution was found in
\cite{Kirsch} which describes the late-time evolution of an expanding 
\mbox{${\cal N}=4$} super Yang-Mills plasma with a single chemical potential.  
Similarly, we now construct a late-time solution from the boosted black
brane solution (\ref{0thsol}) (dual to three chemical potentials).
Proceeding as in \cite{Kirsch}, we
assume the late-time behavior
$m = \tilde\tau^{-4/3} m_0$, 
$q^a = \tilde\tau^{-2/3} q^a_0$ 
for the parameters $m$ and $q^a$ 
and find the zeroth-order solution ($v=\tilde \tau^{1/3} r$)
\begin{align}
ds^2 &= -H^{-{2\over 3}}(v)f(v) d\tilde\tau^2 
+ 2 H^{-{1\over 6}}(v) d\ttau  dr \nonumber \\
&~~~+  H^{1\over 3}(v) \left( (1+r\ttau)^2 dy^2
+ r^2 dx_\perp^2 \right) \,,\nonumber\\
A^a &= - A_0^a(v) d\ttau  +  {\cal A}^a_\mu  dx^\mu \,,\quad
X^a \,\,=\,\, {H^{1\over 3}(v) \over H_a(v)}\,, \label{0thsol2}\nonumber\\
f(v)&=r^2\left(-{m_0\over v^4}+ H(v)\right)\,,\quad H(v)=\prod_{a=1}^3 H^a(v)\,,\nonumber\\
 H^a(v)&=1+{q_0^a \over v^2}\,,\qquad A_0^a(v) = \frac{1}{\tilde \tau^{1/3}} \frac{\sqrt{m_0 q_0^a}}{v^2+q_0^a}  \,.
\end{align}
This background is a good approximation of the full time-dependent 
solution at large $\ttau$ (as we have explicitly checked using computer algebra for 
${\cal A}_\mu^{1} =0$, ${\cal A}_\mu^{2,3} = (0,-x_2 B,0,0)$).
At smaller $\ttau$, it receives subleading corrections in $\ttau^{-2/3}$ corresponding to higher-order gradient corrections.
It has been shown many times that first-order transport coefficients
appear in the first correction in $\ttau^{-2/3}$.
It therefore follows from the above discussion 
that the conductivities $\xi^{ab}_B$,
relevant for the CME, appear in the first-order correction to the solution~(\ref{0thsol2}). 

We are grateful to Karl Landsteiner and Ho-Ung Yee for useful comments and
Email correspondence.

\end{document}